\documentclass[aps,twocolumn,amsmath,amssymb,floatfix,pra,reprint,showpacs,footinbib,superscriptaddress,longbibliography]{revtex4-1}

\usepackage{epsfig,graphicx}
\usepackage{setspace}
\usepackage[english]{babel}
\usepackage{amsfonts}
\usepackage{amsmath}
\usepackage{latexsym}
\usepackage{graphics,bm}
\usepackage{natbib}
\usepackage{dcolumn}
\usepackage{bm}
\usepackage{rotating}
\usepackage{amsmath}
\usepackage{epstopdf}

\usepackage{color}

\usepackage{soul,xcolor}
\setstcolor{red}

\usepackage[T1]{fontenc}
\usepackage[applemac]{inputenc}
\usepackage{lmodern}

\usepackage{ae}
\usepackage{units}

\usepackage{color}
\usepackage{url}

\usepackage[colorlinks]{hyperref}
\hypersetup{%
        plainpages=true,
        breaklinks=true,% not default in dvips mode, so we must specify
        hypertexnames=false,%not ideal, but needed when pagenums duplicate (`i' vs. `1')
        pageanchor=true,
        colorlinks=true,
        linkcolor={blue},
        citecolor={red},
        urlcolor={blue},
        anchorcolor={black}
      }

\newcommand{\ex}[1]{\langle #1 \rangle}

\newcommand{\beq}{\begin{eqnarray}}
\newcommand{\eeq}{\end{eqnarray}}
\newcommand{\ket}[1]{|#1\rangle}

\newcommand{\ketbra}[2]{\left| #1 \rangle \langle #2 \right|}
\newcommand{\brakket}[3]{\left\langle #1\left| #2 \right| #3\right\rangle}
\newcommand{\expec}[1]{\left\langle #1 \right\rangle}
\newcommand{\tr}[1]{\text{tr}\left( #1 \right)}

\newcommand{\abssq}[1]{\left| #1 \right|^2}

\newcommand{\nn}{\nonumber}

\newcommand{\figref}[1]{\mbox{Fig.~\ref{#1}}}

\newcommand{\secref}[1]{\mbox{Sec.~\ref{#1}}}

\newcommand{\appref}[1]{\mbox{Appendix~\ref{#1}}}
\renewcommand{\eqref}[1]{\mbox{Eq.~(\ref{#1})}}

\newcommand{\be}{\begin{equation}}
\newcommand{\ee}{\end{equation}}
\newcommand{\bea}{\begin{eqnarray}}
\newcommand{\eea}{\end{eqnarray}}

\begin{document}

\title{Leggett--Garg inequality violations with a large ensemble of qubits}

\author{Neill Lambert}
\email[e-mail:]{nwlambert@gmail.com}
\affiliation{CEMS, RIKEN, Wako-shi, Saitama 351-0198, Japan}

\author{Kamanasish Debnath}
\affiliation{CEMS, RIKEN, Wako-shi, Saitama 351-0198, Japan}
\affiliation{Amity Institute of Applied Sciences, Amity University, Noida - 201303 (U.P.), India}

\author{Anton \surname{Frisk Kockum}}
\affiliation{CEMS, RIKEN, Wako-shi, Saitama 351-0198, Japan}

\author{George C. Knee}
\affiliation{NTT Basic Research Laboratories, NTT Corporation, 3-1 Morinosato Wakamiya, Atsugi, Kanagawa 243-0198, Japan}
\affiliation{Department of Materials, University of Oxford, Parks Road, Oxford OX1 3PH, United Kingdom}
\affiliation{Department of Physics, University of Warwick, Gibbet Hill Road, Coventry CV4 7AL, United Kingdom}

\author{William J. Munro}
\affiliation{NTT Basic Research Laboratories, NTT Corporation, 3-1 Morinosato Wakamiya, Atsugi, Kanagawa 243-0198, Japan}

\author{Franco Nori}
\affiliation{CEMS, RIKEN, Wako-shi, Saitama 351-0198, Japan}
\affiliation{Department of Physics, University of Michigan, Ann Arbor, Michigan 48109-1040, USA}

%\address{$^1$CEMS, RIKEN, Wako-shi, Saitama 351- 0198, Japan \\
%$^2$Institute of Applied Sciences, Amity University, Noida - 201303 (U.P.), India \\
%$^3$Department of Materials, University of Oxford, Parks Road, Oxford OX1 3PH, UK \\
%$^4$NTT Basic Research Laboratories, NTT Corporation, 3-1 Morinosato Wakamiya, Atsugi, Kanagawa 243-0198, Japan \\
%$^5$Department of Physics, University of Michigan, Ann Arbor, Michigan 48109-1040, USA}

\date{\today}

\begin{abstract}
We investigate how discrete internal degrees of freedom in a quasi-macroscopic system affect the violation of the Leggett--Garg inequality, a test of macroscopic-realism based on temporal correlation functions. As a specific example, we focus on an ensemble of qubits subject to collective and individual noise. This generic model can describe a  range of physical systems, including atoms in cavities,  electron or nuclear spins in NV centers in diamond, erbium in Y$_2$SiO$_5$, bismuth impurities in silicon, or arrays of superconducting circuits, to indicate but a few. Such large ensembles are potentially more macroscopic than other systems that have been used so far for testing the Leggett--Garg inequality, and open a route toward probing the boundaries of quantum mechanics at macroscopic scales. We find that, because of the non-trivial internal structure of such an ensemble, the behavior of different measurement schemes, under the influence of noise, can be surprising.  We discuss which measurement schemes are optimal for flux qubits and NV centers, and some of the technological constraints and difficulties for observing such violations with present-day experiments.
\end{abstract}

\pacs{03.65.Ta, 85.25.Cp}

% 03.67.Mn 	Entanglement measures, witnesses, and other characterizations
% 03.65.Ta 	Foundations of quantum mechanics; measurement theory
% 42.50.Dv 	Quantum state engineering and measurements
% 42.50.Pq 	Cavity quantum electrodynamics; micromasers
% 85.25.Cp 	Josephson devices

\maketitle

\section{Introduction} The crossover between classical and quantum worlds still remains under debate, even 80 years after  Schr\"odinger's famous `cat' thought experiment \cite{Schrodinger1935}. For example, the precise details of how the classical macroscopic world arises from the quantum one, and whether there is an unknown fundamental boundary between the two, still remains a topic of vigorous study. In 1964, Bell \cite{Bell1964}  made the assumptions of \textit{realism} and \textit{locality} to derive an inequality for correlations between spatially separated events, whose violation can rule out certain classes of alternative theories to quantum mechanics.  More recently, Leggett and Garg \cite{Leggett1985} asked a related but different question: can a large, macroscopic, system be in a genuine quantum superposition, or is there some unknown mass, particle number, or length scale limit where substantial corrections to quantum theory prevent such a state of affairs? To give a quantitative tool to test for such breakdowns they assumed the twin assumptions of \textit{`macroscopic realism'} and \textit{`noninvasive measurability'} to construct what is now known as the Leggett--Garg inequality (LGI) \cite{Leggett1985, Emary2014}. Violations of this inequality by large systems rules out certain classes of non-invasive realistic theories (henceforth termed macrorealism), and provide evidence of quantum effects at the macroscopic scale. With advancements in fabrication techniques \cite{Stern2014, You2011,Chow2010} and a number of LGI violations being reported in microscopic systems~\cite{Goggin2011,KneeSimmonsGauger2012,Palacios-LMalletNguyen2010,Robens2015}, it has become important to test the inequality on arguably `larger' macroscopic systems~\cite{GYSciRep}, and push back further the demarcation between quantum and classical worlds.

\begin{figure*}
\centering
\includegraphics[width=\textwidth]{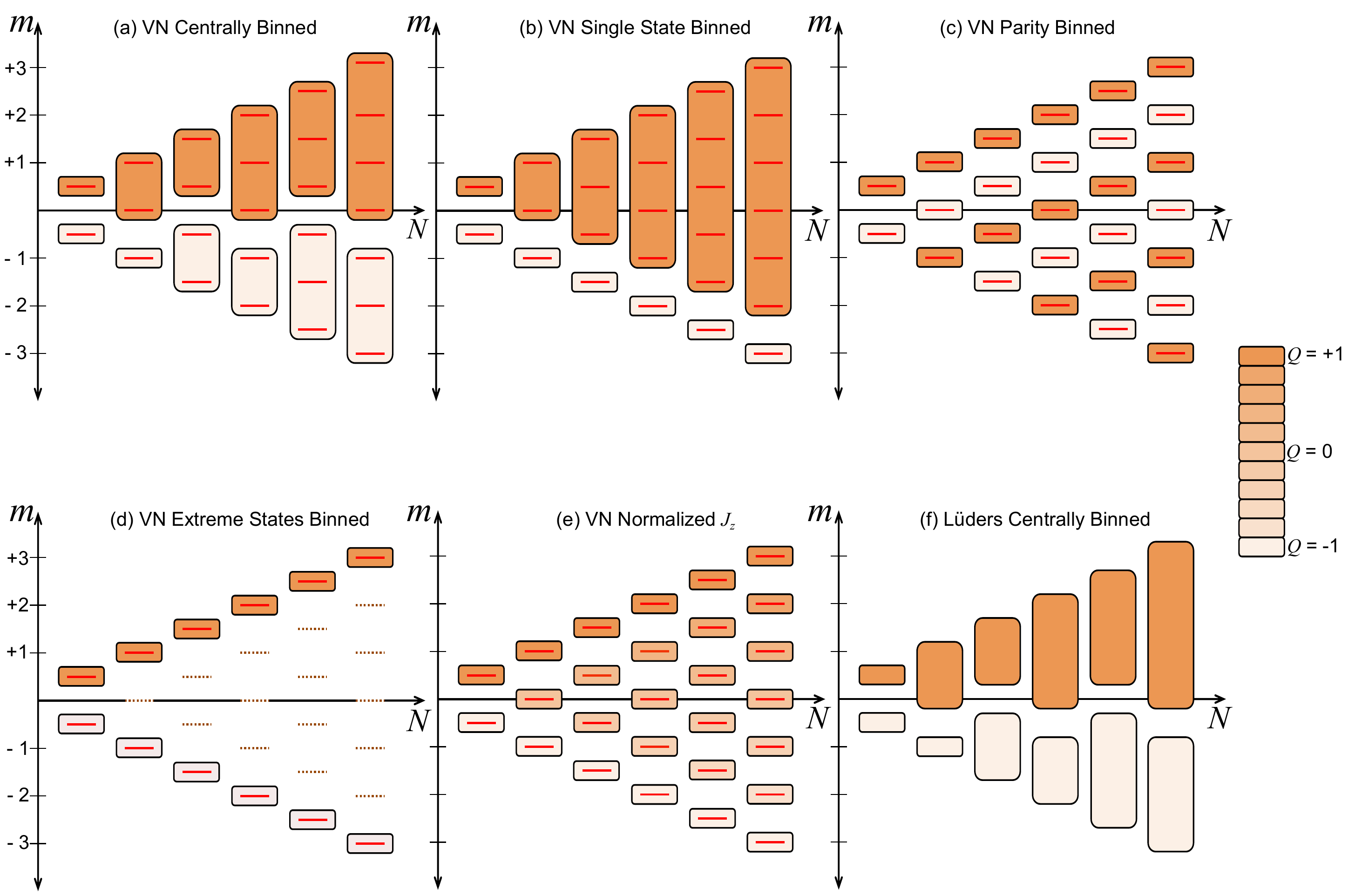}
\caption{(Color online) The six different measurement schemes at a glance. Note that the ordering of the schemes we use in this figure is replicated in Figs.~\ref{fig:KvsTime} -- \ref{grid}. The boxes enclose levels giving the same value for $Q$ and the box coloring corresponds to the $Q$ value. Solid lines for levels within the boxes indicate that the measurement projects the system onto that specific state. Measurement results for levels not enclosed by boxes are discarded. a) VN Centrally Binned: $q_{m>0}=+1$ and $q_{m \leq 0}=-1$. b) VN Single State Binned: $q_{-j}=-1$, and $q_{m>-j}=+1$. c) VN Parity Binned: $q_m = +1$ for $m = j, j-2, j-4, \ldots$ and $q_m = -1$ for $m = j-1, j-3, j-5, \ldots$. d) VN Extreme States Binned: $q_j = +1$, $q_{-j} = -1$, and all other measurement results are discarded. e) VN Normalized $J_z$ measurement: $q_m = m /j$. f) L\"uders Centrally Binned: $q_{m>0}=+1$ and $q_{m \leq 0}=-1$, but, unlike the other schemes, the measurement does not project further within these two subspaces.}
%The normalized $J_z$ scheme (green) relies on measuring $J_z$ and normalizing the final measurement result. The dichotomic measurement schemes (red), both the von-Neumann and Luder form,  return just $\pm 1$ depending on a pre-selected binning of the whole state space (here it is divided around $m=0$).  These two schemes differ in whether the measurement device resolves the sub-levels within the binned space. Finally,  the extreme states measurement scheme (blue) records only results where the system is found in one of the two extremal states, and discards all other results.}
\label{fig:schemes}
\end{figure*}

Alongside this lingering fundamental question, advancements in nanomechanical devices \cite{Harris1996} such as suspended resonators \cite{Treutlein2007}, opto-mechanical mirrors \cite{Sankey2010,Lambert2011} and vibrating membranes \cite{Jayich2008}, have generated interest in understanding the crossover from the quantum to classical regimes. Similarly, circuit QED \cite{You2011, Xiang2013} has helped in the exploration of phenomena such as superradiance \cite{Viehmann2011} and entanglement \cite{Chow2010} in low-noise environments \cite{Eichler2013} with quasi-macroscopic systems.  In addition, it is becoming apparent that the physics of systems with internal structure, which cannot be assumed to be restricted to a simple two-level Hilbert space, is both rich and useful; Budroni and Emary \cite{Budroni2014} found that the magnitude of the violation can increase as the number of internal levels increases, reaching an upper bound, a temporal analogue to the Tsirelson bound \cite{tsirel} for the Bell Inequality. Additionally, George {\em et al.}~\cite{George2013}  found that a multi-level system could exhibit a violation of the LGI while not violating a related condition  known both as the quantum witness equality~\cite{Li2012}, or `no-signalling in time' \cite{Kofler2013} --- arguably allowing one to discount a stricter class of macrorealist theories.  Finally, Budroni {\em et al.}~\cite{Budroni15} considered the continuum limit of a macroscopic ensemble, and characterized the requirements on measurements in this case.

Here, we theoretically investigate the LGI in a discrete ensemble of $N$ two-level quantum systems, physical manifestations of which include arrays of nearly identical flux qubits, NV centers in diamond, erbium in YSO and bismuth impurities in silicon~\footnote{These examples comprise a rich energy level structure, and we assume here that two levels have been isolated well from the rest (for example by applying a magnetic field), so that they are the only relevant states.}. In particular, we shall study the effects of various choices of measurement schemes, from the point of view both of the degree of macroscopicity and the feasibility of observing a LGI violation. %As we show in detail, some of the above mentioned physical realizations may be electromagnetically coupled to a microwave transmission line cavity, or a large SQUID \cite{Matsuzaki2015} for this purpose.

We select and investigate six different measurement schemes, which are all defined in the fully symmetric subspace of the $N$ qubits. This subspace forms a ladder of state manifolds indexed by the number $n$ of excited qubits in that manifold $n\in\{0,1,2\ldots N\}$, and can thus be viewed as an $(N+1)$-dimensional system which we call the `large spin'. Following convention, we use the simply-related variable $m=n-N/2$. Setting $j=N/2$, our ladder is indexed by the label $m\in\{-j,-j+1,\ldots,j-1,j\}$. In the noise-free case, we evaluate, where feasible, both analytic expressions and numerical simulations of the Leggett-Garg parameter and attempt to extrapolate large-$N$ limits.  We then consider, numerically, each scheme's performance  under both collective and individual qubit noise.  Finally, we consider whether each scheme allows for a macroscopic interpretation of an observed violation.

Based on the above analysis, our main results in this work are two-fold.  Firstly, among the options considered here, a measurement which  distinguishes sub-states of the collective large-spin Hilbert space, and which bins around the center of that that space, gives a violation of the LGI which is the most robust against noise.  In addition, this violation does not vanish as $N \rightarrow \infty$.  Secondly, in contrast, we find that if one wishes to fully explore the notion of macroscopic quantum effects in such systems, a measurement which only returns information on extreme states of the collective large-spin Hilbert space is the most ideal.  However, while robust against dephasing, such a measurement is sensitive to both collective and individual dissipation, and violations may be masked by such unwanted noise.

\section{The Leggett--Garg Inequality}
We consider the  Leggett-Garg parameter in the form
\begin{equation}
K =C_{21}+ C_{32}- C_{31},
\label{eq:DefK}
\end{equation}
where $C_{\beta\alpha}$ is the correlation function of a dichotomic variable $Q=\pm 1$ measured at two times, $t_{\beta} > t_{\alpha}$, such that $C_{\beta\alpha}\equiv \ex{Q(t_{\beta})Q(t_{\alpha})}$~\cite{Leggett1985,Emary2014}.  Leggett and Garg derived~\cite{Leggett1985}, under the assumptions of macroscopic realism and non-invasive measurement, their inequality $K\leq 1$, and showed that a quantum two-level system easily violates this bound. While, as with the Bell Inequality, many forms of the inequality exist\cite{Lambert2010,Emary2012, Emary2014} we employ this form as it is typically violated for short time-intervals between measurements.  In addition, although the LGI is not a sufficient condition for macrorealism (unlike the related condition derived in \cite{Li2012,Kofler2013}),  the LGI remains nevertheless a necessary condition whose violation implies the failure of at least one of Leggett and Garg's assumptions~\cite{Leggett1985}. Furthermore, the LGI has various attractive properties~\cite{Emary2014} not shared by other conditions --- for example, it is possible to find state-independent violations \cite{Fritz10},  allowing use of the highly mixed thermal states we expect to describe some qubit ensembles.

The spirit of the LGI is to perform experiments on larger and larger systems, checking for a violation of this, or an equivalent, inequality (having removed all sources of decoherence and dissipation that one can control and understand from within quantum mechanics itself). A violation would then rule out macrorealism at that scale. A macrorealist might argue either (i) that there are broader classes of alternative theories to quantum mechanics, particularly ones which include invasive measurements in a fundamental way or (ii) that the violation is due simply to clumsy measurements.  One way to go beyond such doubts is to combine the LGI with a test of how invasive the measurements themselves are \cite{Halliwell2015, George2013, Knee2016}. Such an analysis in the context of large ensembles  would be a fruitful topic of future research.

\begin{figure*}
\includegraphics[width=1\textwidth]{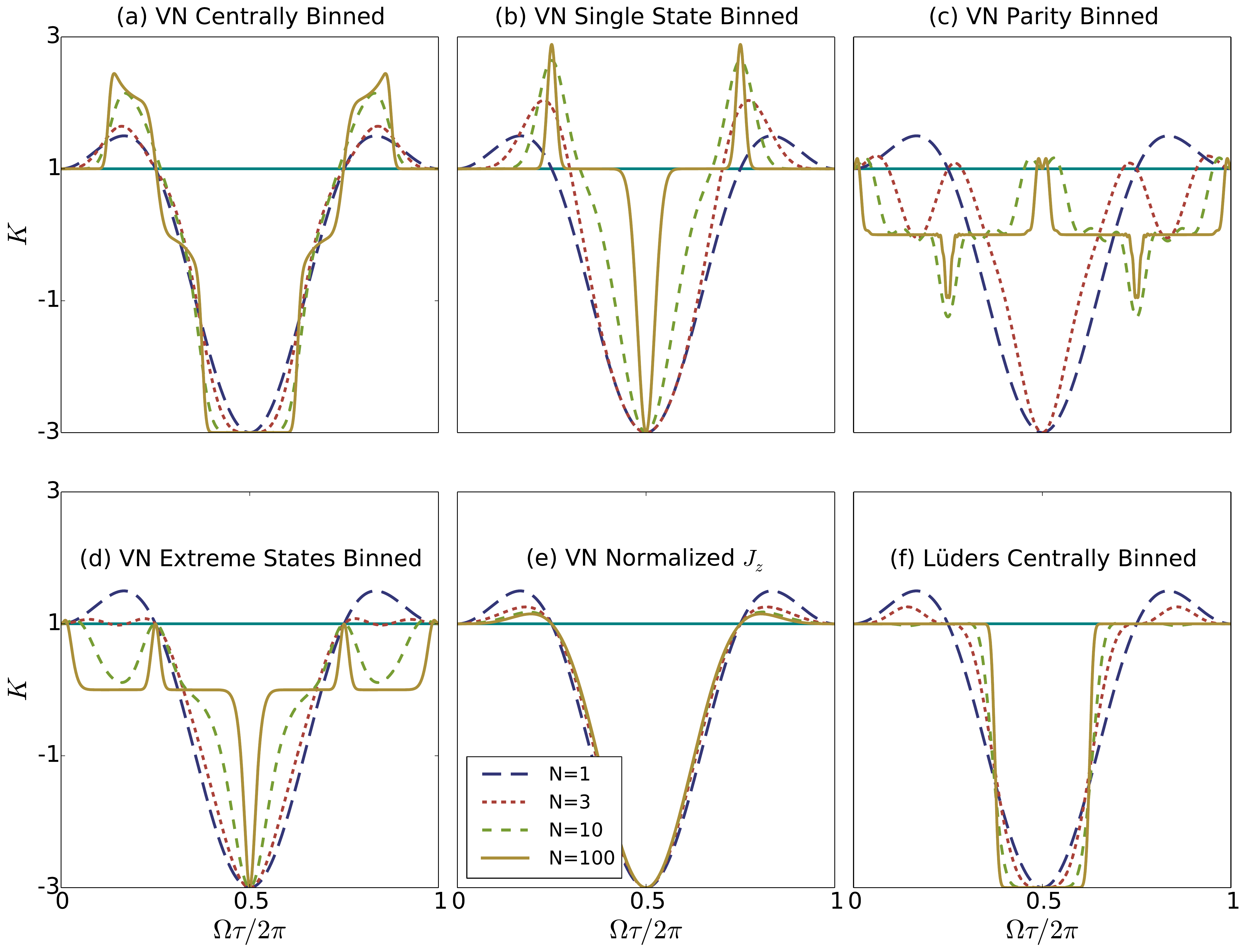}
\caption{(Color online) Variation of the LG parameter $K$ as a function of time for all six schemes (see \figref{fig:schemes} for a schematic explanation of each scheme) with $N = \{1, 3, 10, 100\}$.  The turquoise line in each figure marks the classical bound $K=1$. Note that all schemes converge to the same result for a single qubit, $N=1$.}
\label{fig:KvsTime}
\end{figure*}

\section{Model \& measurement schemes}

We will find it useful to define $\sigma_{x}^{(k)}$ and $\sigma_{z}^{(k)}$ as the Pauli $x$ and $z$ matrices (respectively) for qubit $k=1,2,...,N$. We then consider the dynamics governed by the Hamiltonian
\begin{equation}
H= \hbar \frac{\omega_q}{2}\sum_{k=1}^N \sigma_z^{(k)} + 2 \hbar \Omega \cos(\omega_d t) \sum_{k=1}^N \sigma_x^{(k)},
\label{eq:Hamiltonian}
\end{equation}
where $\omega_q$ is the energy splitting of the qubits, which we assume to be homogenous, and $\Omega$ is a transverse drive. This allows us to use standard spin-resonance techniques to obtain the effective Hamiltonian in the interaction picture, and under the rotating-wave approximation, so that when $\omega_d = \omega_q$ and $\Omega \ll \omega_q$, the ensemble Hamiltonian becomes
\begin{equation}
H^{{\mathrm{(I)}}}= \hbar \Omega \sum_{k=1}^N \sigma_x^{(k)} \equiv  \hbar \Omega J_x.
\label{eq:Hamiltonian2}
\end{equation}
Here we have used the collective operator $J_x=\sum_{k=1}^N  \sigma_x^{(k)}$, which represents the $x$ component of the angular momentum operator defining the ensemble behavior of the $N$ qubits. We also define the collective lowering operator as $J_-=\sum_{k=1}^N \sigma_-^{(k)}$, and the $z$ component of the angular momentum operator as $J_z=\sum_{k=1}^N \sigma_z^{(k)}/2$.  In all of the following we operate in the interaction picture, and drop the label $\mathrm{(I)}$  from the Hamiltonian.

To find a violation of the LGI, we fix the initial state of our $N$ qubits to the fully polarized state $\psi (t=0) = \ket{\!\!\!\uparrow \uparrow \ldots \uparrow}$ in the $z$ direction. Note that, in the pure evolution case, the results are largely independent of the initial condition. However, in the presence of noise, particularly dissipation in the $z$ basis, this initial condition is favourable to give large violations for large $N$ for most schemes.  In terms of the collective operators, one has $J_z \psi (t= 0)= j \psi (t= 0)$, i.e., the initial state is the highest weight $m=j$ eigenstate of our large spin in the $z$-direction. In constructing the correlation functions used in the LGI, we assume that we perform measurements in the $z$ basis at consecutive times $t_1=0$, $t_2=\tau$, and $t_3=2\tau$. The $z$ basis is chosen to be the one which couples to the measurement device and thus corresponds to a physical observable of the macroscopic ensemble.

In addition to the above unitary evolution, we also assume that each qubit can experience individual dephasing, with rate $\gamma_D$, and dissipation $\gamma_L$. In addition, we assume that the ensemble as a whole can experience a collective dephasing $\Gamma_D$ and dissipation $\Gamma_L$. These act on the individual or collective $z$ basis, as this is the fundamental energy basis of our ensemble in the lab frame. The total dynamics is then described by the master equation,
\bea
\dot{\rho} &=& \mathcal{M}[\rho]=-\frac{i}{\hbar}[H,\rho]  \nn \\
&&+ \sum_{k=1}^N\left\{ \frac{\gamma_D}{2} \mathcal{L}\left[\sigma_z^{(k)}\right]\rho+ \gamma_L \mathcal{L}\left[\sigma_-^{(k)}\right]\rho \right\} \nn \\
&&+ 2\Gamma_D \mathcal{L}\left[J_z\right]\rho+ \Gamma_L \mathcal{L}\left[J_-\right]\rho,
\label{eq:MasterEq}
\eea
where $\rho$ is the density matrix of the system, $\mathcal{L}[a]$ is the Lindblad operator $\mathcal{L}[a]\rho=a \rho a^\dag - \frac{1}{2}\{a^\dag a, \rho\}$, and we assume negligible temperature. Note that we have scaled $\Gamma_D$ so that the collective and individual dephasing contributions are equivalent in the $N=1$ limit.  When all the dephasing and dissipation terms are zero, we can often obtain analytical results, as will be described below, and in detail in appendix \ref{AppA}. When the collective dephasing or dissipation are non-zero, $\Gamma_i\neq 0$, we numerically solve \cite{qutip1,qutip2} the above master equation within the large spin $(N+1)$-dimensional restricted Hilbert space.  When the individual qubit dephasing or dissipation is non-zero, $\gamma_i \neq 0$, we perform numerical simulations which take into account the full $2^N$ Hilbert space of the ensemble. This restricts us to investigating a smaller range of $N$ (due to having only finite computational resources).

Our measurement schemes fall into two classes, depending on the physics of the measurement process itself. The first class relates to a projective measurement of $J_z$, followed by one of five different data-processing steps or `binning' strategies. Immediately after the measurement, the ensemble is left in an eigenstate of $J_z$: the appropriate state update rule is that of von Neumann (VN)
\be
\rho^M \rightarrow \sum_{m}  q_m  \ketbra{m}{m} \rho  \ketbra{m}{m},
\label{eq:UpdateRule}
\ee
where $\rho$ is the state immediately before the measurement, and $\rho^M$ is the state immediately after. The data-processing step, however, `compresses' the eigenvalue and reduces it to $\pm1$ according to one of a set of predetermined rules (introduced below). The second class of measurement relates to a projective measurement of a different observable, where each projector is a sum of $J_z$ eigenprojectors. Because the binning is performed \emph{prior} to the measurement itself, the system is left in an incoherent mixture of the two binning sub-spaces afterwards. We discuss this further below, in Sec. \ref{Luder}.

The choice of binning strategy and state-update rule allow for a large  number of measurement strategies. This set of strategies was analyzed for the largest possible violation, in the pure-evolution case, in \cite{Budroni2014}, using convex optimization techniques. There, they found that using the VN update rule, and binning the measurement results in terms of a single state versus all others, gave the largest possible violation. Here we instead look in detail at six  distinct, but experimentally motivated, strategies (shown in \figref{fig:schemes}), and how they behave under the influence of noise. %We find that the single state binning is not optimal in the presence of noise. #We conclude this work with details of possible physical implementations of measurement devices, and how the different schemes may correspond to different types of couplings to such devices.

We will begin with the first class of measurement scheme:
given the VN state update rule we can write down an explicit formula for the correlation functions with which we construct the LGI,
\begin{eqnarray}
\expec{Q(t_2) Q(t_1)}= \text{tr} \Bigg[ \sum_k q_k \ketbra{k}{k} \mathcal{U}(t_2- t_1) \nonumber \\
\sum_m q_m \ketbra{m}{m} \rho(t_1) \ketbra{m}{m} \Bigg],
\label{eq:CorrelationFunction}
\end{eqnarray}
where $\mathcal{U}(t_2- t_1)=\exp[\mathcal{M}(t_2-t_1)]$ is the propagator in superoperator form, such that it acts on all operators to the right.

\begin{figure*}[]
\includegraphics[width=\textwidth]{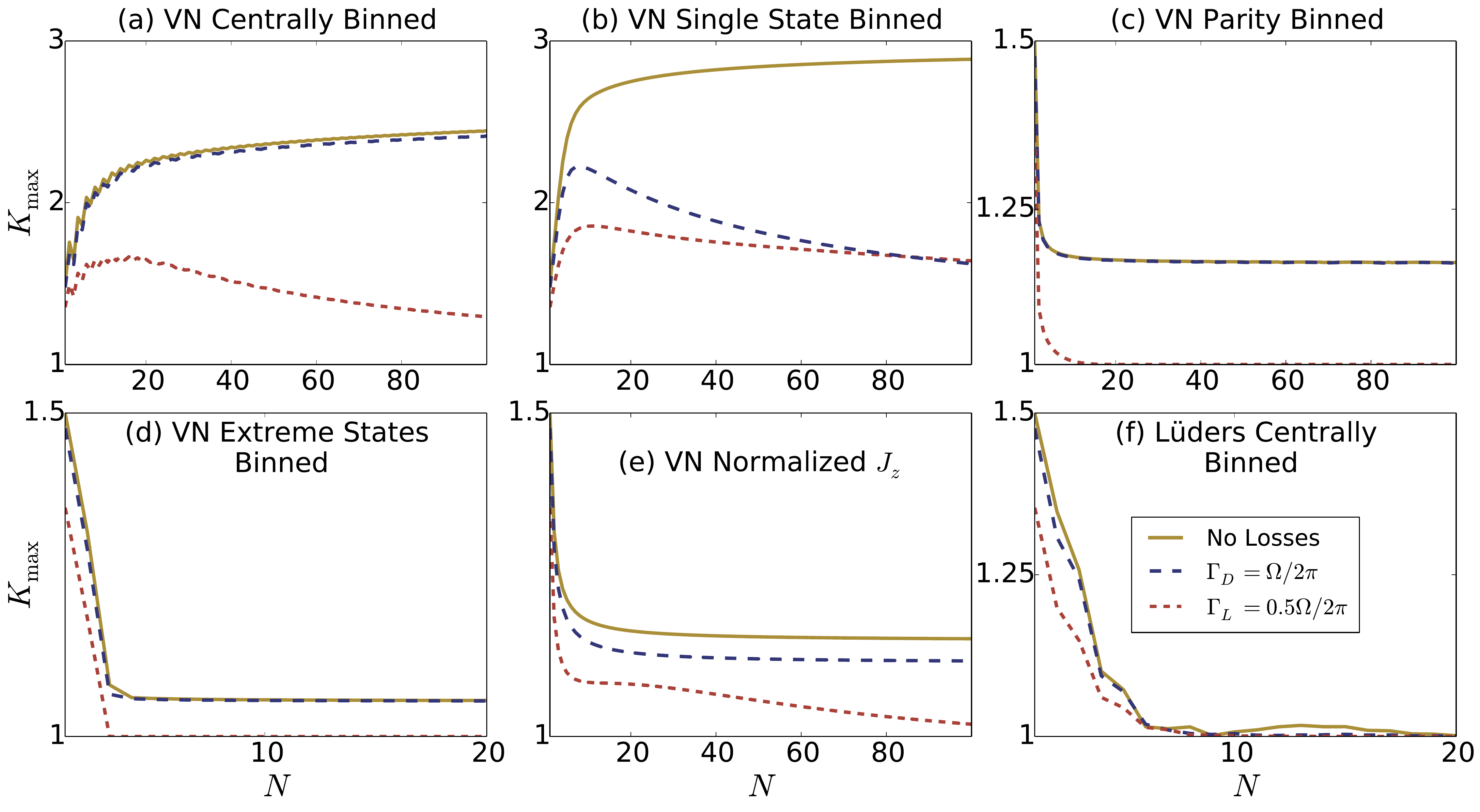}
\caption{(Color online) $K_{\mathrm{max}}$ as a function of $N$, both for the noise-less evolution and in the presence of collective dephasing $\Gamma_D=\frac{\Omega}{2 \pi}$ and collective relaxation $\Gamma_L=\frac{0.5\Omega}{2 \pi}$, for all measurement schemes. The cases with noise were evaluated numerically, but are still amenable to large-$N$ evaluation due to the reduced Hilbert space of a collective spin. In figures (d) and (f) we truncate the x-axis at smaller values of $N$ as both saturate for large $N$ and have interesting features at small $N$ values.}
\label{fig:KmaxvsNCollective}
\centering
\end{figure*}

\begin{figure*}[]
\includegraphics[width=\textwidth]{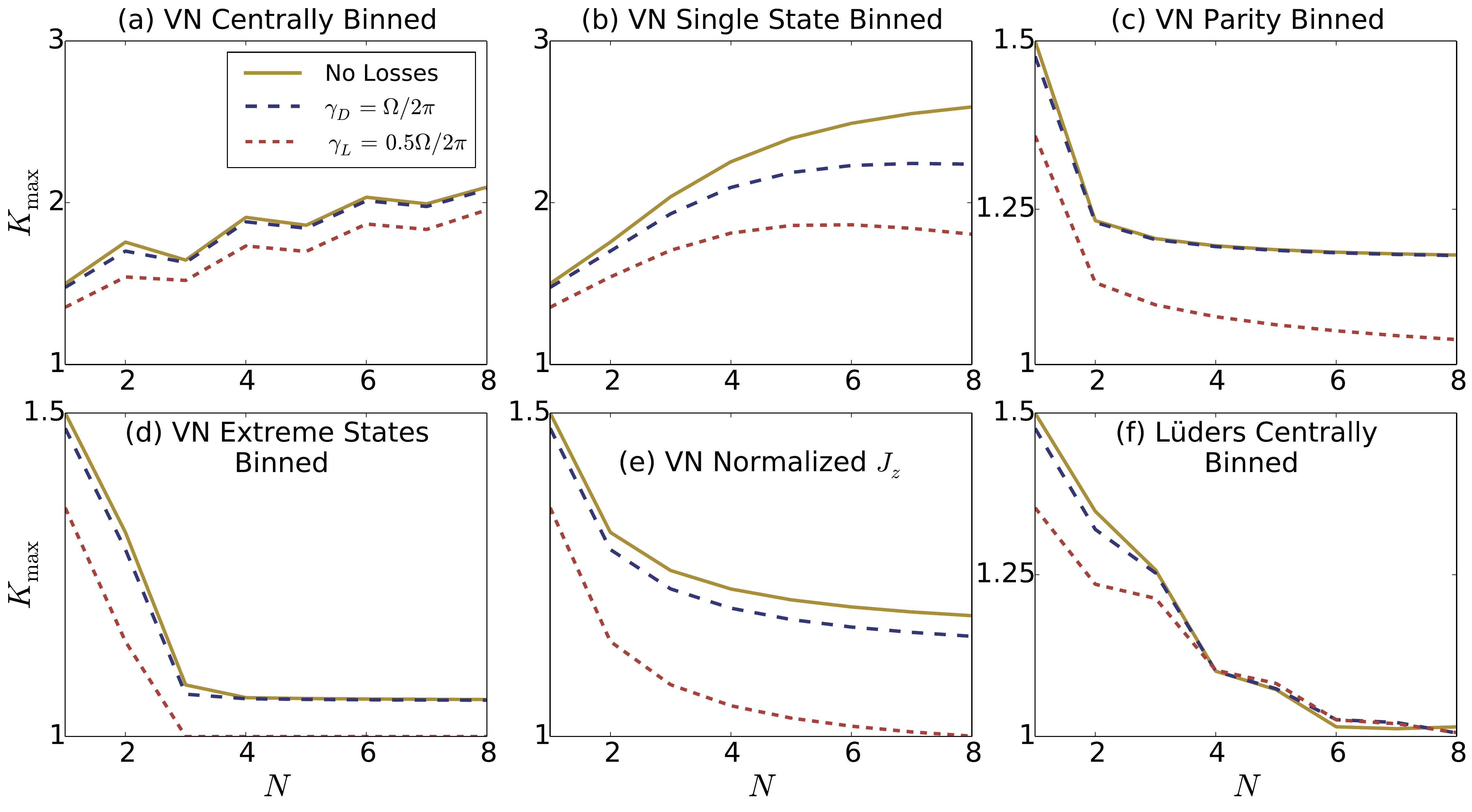}
\caption{(Color online) $K_{\mathrm{max}}$ as a function of $N$, both for case of no noise and in the presence of individual dephasing $\gamma_D=\frac{\Omega}{2 \pi}$ and individual relaxation $\gamma_L=\frac{0.5\Omega}{2 \pi}$ noises. These values are large compared to the noises achieved with currently available flux qubits and NV centers, in order to show an extreme limit.  The results are evaluated numerically, but we are restricted to much smaller values of $N$, as we must include the full $2^N$ Hilbert space for the calculations with such individual decoherence.}
\label{fig:KmaxvsNIndividual}
\centering
\end{figure*}

\subsection{von Neumann centrally binned}
We first consider a binning strategy where, in the above formula, we choose $q_{m\geq0}=+1$ and $q_{m < 0}=-1$.  This choice is depicted schematically in \figref{fig:schemes}a, while \figref{fig:KvsTime}a shows the corresponding LG parameter, $K$, as a function of time for different ensemble sizes. One immediately sees that, for this scheme, the maximum violation increases with the ensemble size, seeming to tend to a maximum around $\Omega \tau= \pi/4$. This dependence of the maxima on the ensemble size is shown more explicitly in \figref{fig:KmaxvsNCollective}a.  In \appref{AppA}, we show how the pure-state results can be calculated analytically.

Within \figref{fig:KmaxvsNCollective}a we show the influence of strong collective noise $\Gamma_D=\frac{\Omega}{2 \pi}$ (dashed line, with $\Gamma_L=0$) and $\Gamma_L=\frac{0.5\Omega}{2 \pi}$ (dotted line, with $\Gamma_D=0$).  The maximum is almost unaffected by the strong collective dephasing  $\Gamma_D$, but is strongly influenced by the collective dissipation  $\Gamma_L$. In \figref{fig:KmaxvsNIndividual}a we show the effect of individual noise for a smaller range of $N$ (due to the drastically larger Hilbert space required, as individual noise breaks the large-spin symmetry, necessitating a full $2^N$ simulation).  Here we see that for this scheme, collective and individual {\em dephasing} have a similar minor effect, while collective {\em dissipation} is much more damaging, for large $N$, than individual noise. The latter can be attributed to collective superradiance  \cite{DickeSR,Gross82} that occurs when a large ensemble of identical emitters experiences collective dissipation. (Note that the equivalent binning scheme $q_{m \geq 0}=+1$ and $q_{m<0}=-1$, which we have not explicitly shown, for odd values of $N+1$ one sees slightly different small-$N$ behavior, but the same large-$N$ limit).

%\emph{(b) VN Single State Binning}
\subsection{von Neumann single state binning}
In Ref.~\cite{Budroni2014} they found that, at least in the closed system case, the largest violation occurred for the choice of $q_{-j}=-1$, and $q_{m>-j}=+1$, i.e., where only one state (the lowest-lying one in the large-spin bases, for example) contributes to one of the binned results, and all the other states to the other binning outcome.  This is shown schematically in \figref{fig:schemes}b, and the time dependence of $K$ is illustrated in \figref{fig:KvsTime}b. As shown in \figref{fig:KvsTime}b, as $N$ is increased one sees an asymptotic limit (for the pure evolution case) that can be evaluated analytically: $K_{\mathrm{max}}(N\rightarrow \infty) \rightarrow 3$  (see \cite{Budroni2014} and our \appref{AppA}).

However, this scheme is sensitive to both collective and individual noise. Figures~\ref{fig:KmaxvsNCollective}b and \ref{fig:KmaxvsNIndividual}b show that, as $N$ increases, the bound is substantially reduced when compared to the pure evolution case.  Unlike the previous case (VN centrally binned) it is quite sensitive to both collective and individual dephasing and \figref{fig:KmaxvsNIndividual}b indicates a crossing where individual noise becomes more detrimental.  One should note that \figref{fig:KvsTime}b suggests the time-window of observing the violation narrows as $N$ increases. This can be attributed to the fact that the dynamics of the system mean the probability of it being in the $q_{-j}=-1$ binning subspace diminishes as $N$ increases. One may hypothesise that this influences the sensitivity to noise we observe in Figs.~\ref{fig:KmaxvsNCollective}b and \ref{fig:KmaxvsNIndividual}b.

%\emph{(c) Parity Binning}
\subsection{von Neumann parity binning}
Another binning strategy previously employed elsewhere \cite{Budroni2014} is to assign the $Q$ values according to the parity of the $J_z$ states.  As an example here, we use $q_m = +1$ for $m = j, j-2, j-4, \ldots$ and $q_m = -1$ for $m = j-1, j-3, j-5, \ldots$(see \figref{fig:schemes}c).  One immediately notices, in \figref{fig:KvsTime}c, that the maximum violation diminishes as $N$ increases, apparently reaching a small constant value with an initial maxima at small times. %Again, one can derive an analytical form for the time dependance, but it is not amenable for extracting an asymptote. However, as with the centrally binned example (a), we can evaluate these expressions for large N and find  $K_{max}(N\rightarrow \infty) \rightarrow ...$.
The behavior under collective dephasing appears robust, but collective dissipation (\figref{fig:KmaxvsNCollective}c) has a strong influence even at moderate $N$ values, entirely removing the violation.

%\emph{(d) Extreme state measurement--}
\subsection{von Neumann extreme states binning}

Various precise definitions of what constitutes a truly ``macroscopic"  superposition abound. A necessary but not sufficient criterion proposed by Leggett himself was the  ``extensive difference" of the possible measurement results, i.e., difference in the expectation value, normalized to some appropriate atomic scale, between the two dichotomic outcomes.  In the schemes we have discussed so far, even for large $N$, it is difficult to {\em a priori} look at the LG inequality and argue that the violation arises due to the coherence between macroscopically distinct states (\textit{e.g.}, the evolution could, in principle, be constrained to a subspace of states differing only by $\Delta m \ll N$). Given this motivation to make the definition of `macroscopic' more vivid, we are motivated to consider only the extreme sublevels of any $N$ ensemble, namely for measurement results where $m \neq \pm j$ are discarded (assigned $q=0$), while the extreme states are binned according to $q_j = +1, q_{-j} = -1$, as these are the most distinct (see \figref{fig:schemes}d). For a fuller discussion of this notion of macroscopicity, see \secref{macroscopicity}.

For this choice of measurement scheme, \figref{fig:KvsTime}d shows the variation of $K$ with time for different ensemble sizes, and \figref{fig:KmaxvsNCollective}d shows how the maximum changes with $N$. We see that the maximum violation diminishes but saturates at large N, such that even though we throw away many intermediate states, a violation with a large ensemble is still possible (albeit in a shorter and shorter time window, as per schemes (b) and (c)). In \appref{AppA}, we show how to evaluate the noise-free result analytically, which in this case reduces to a manageable form, giving, for the full LGI,
%\begin{widetext}
\beq
K &=& \left[\cos\left(\frac{\Omega\tau}{2}\right)\right]^{4j} - \left[\sin\left(\frac{\Omega\tau}{2}\right)\right]^{4j} \nn\\
&&+ \left[\cos\left(\frac{\Omega\tau}{2}\right)\right]^{8j} - \left[\sin\left(\frac{\Omega\tau}{2}\right)\right]^{8j} \nn\\
&&- [\cos(\Omega\tau)]^{4j} + [\sin(\Omega\tau)]^{4j}.
\label{eq:KExtremeStates}
\eeq
We find that, resolving the LGI for very large values of $N$ suggests $K_{\mathrm{max}} \rightarrow 1.055$.

This binning strategy is, like the parity binning, robust to collective and individual dephasing as $N$ increases, but is sensitive to both collective and individual dissipation (see Figs.~\ref{fig:KmaxvsNCollective}d and \ref{fig:KmaxvsNIndividual}d).  Thus, while physically appealing due to its clearer ``macroscopic" interpretation, this approach represents an experimental challenge in truly large systems.

The possibility of finding a larger violation (with this measurement scheme) by engineering a more complicated dynamics (e.g., a coupling between just the extreme states) for the ensemble would be an interesting line of future enquiry.
%\emph{(e) Normalized $J_z$ measurement--}
\subsection{von Neumann normalized $J_z$ measurement}

The LGI allows for not just truly dichotomic outcomes, but also for normalized expectation values. As long as these values are bounded, one can derive the LGI without any loss of generality. For completeness, here we show how taking this approach influences the violation.  Again, we assume that our measurement device can distinguish the $(N+1)$ eigenstates $m$, but that our measurement outcomes are binned in such a way that they correspond to the measurement of the large spin $J_z$ operator, normalized by $N/2$. In other words,  $q_m = m /j$ (see \figref{fig:schemes}e). Figure \ref{fig:KvsTime}e shows the variation of $K$ as a function of time for varying ensemble size. The violation diminishes and saturates to a constant value as a function of $N$, as seen in \figref{fig:KmaxvsNCollective}e (again, see \appref{AppA} for an analytical formula for the noise-free case).  The influence of collective noise in this case is once again quite strong, with dephasing reducing the maxima, and dissipation again removing the violation completely for large $N$, though in this case the influence of individual dissipation is more detrimental, as seen in \figref{fig:KmaxvsNIndividual}e.

%\emph{(f) Dichotomic measurement with L\"uder's state update rule--}
\subsection{ L\"uders state update rule\label{Luder} with central binning}

Finally, in contrast to all the previous examples, we consider the case where our measurement device is not capable of distinguishing which of the $m$ sublevels the system is in.  Modelling such a measurement requires a slightly different definition of the post-measurement state. We assume that the measurement device distinguishes the $m \geq 0$ and $m<0$ subspaces, and bins accordingly (illustrated schematically in \figref{fig:schemes}f), thus, following the definition of L\"uders~\cite{Luder06}, the post-measurement state is
\begin{eqnarray}
\rho^m \rightarrow \left( \sum_{m=0}^{m=j} \ketbra{m}{m} \right) \rho \left(\sum_{m= 0}^{m= j} \ketbra{m}{m} \right)-  \nonumber \\
\left( \sum_{m=-j}^{m=-1} \ketbra{m}{m} \right) \rho \left(\sum_{m= -j}^{m= -1} \ketbra{m}{m} \right).
\label{eq:LuderPostMsmt}
\end{eqnarray}
In related works, Brukner \textit{et al.} \cite{Kofler2007,BruknerQW} argued that a similar type of coarse-grained measurement makes the system appear more classical, and termed such a measurement ``fuzzy".  Again, in \figref{fig:KvsTime}f we show the behavior of the LGI for different values of $N$. For $N>20$ the violation disappears, even in the noise-free case, as a direct consequence of the reduced quantum invasiveness of the measurement:  this is a clear illustration that the nature of the LGIs sensitivity is directly related to how invasive the quantum mechanical measurements are on the dynamics of the system.  Note that Budroni {\em et al.}~\cite{Budroni2014} and Fritz \cite{Fritz10} characterize the L\"uders example as being equivalent to a two-level system.  However, here we prepare the ensemble in a state which is not an eigenstate of the subspace binning, and evolve under a Hamiltonian which does not respect the subspace binning of the measurement, leading to a weaker violation as $N$ increases.

In addition, intriguingly, there are two examples of non-monotonic violations with this scheme.  In \figref{fig:KmaxvsNCollective}f we see that the violation has a minimum around $N=8$ and an increase at $N=9$, until decreasing again for larger $N$.  Similarly, in \figref{fig:KmaxvsNIndividual}f we see that between $N=5$ and $N=8$ the maximum of the violation is slightly enhanced, over that seen with the noise-free result, by individual dissipation and dephasing.  Both of these unique features, not seen in other schemes, may be attributable to the fuzzy nature of this measurement; we can only observe violations of the LGI when the state of the system has significant coherence between the $m\geq 0$ subspace and the $m<0$ subspace.  In the presence of noise, while coherence is reduced overall, it is possible for both dephasing and dissipation to induce a faster  evolution towards states near $m=0$, giving rise to the noise-enhanced features we see here.  In future work it may be useful to explore this feature further, and see if similar features arise in the quantum-witness form of the LGI inequality.
\begin{figure*}
\includegraphics[width=\textwidth]{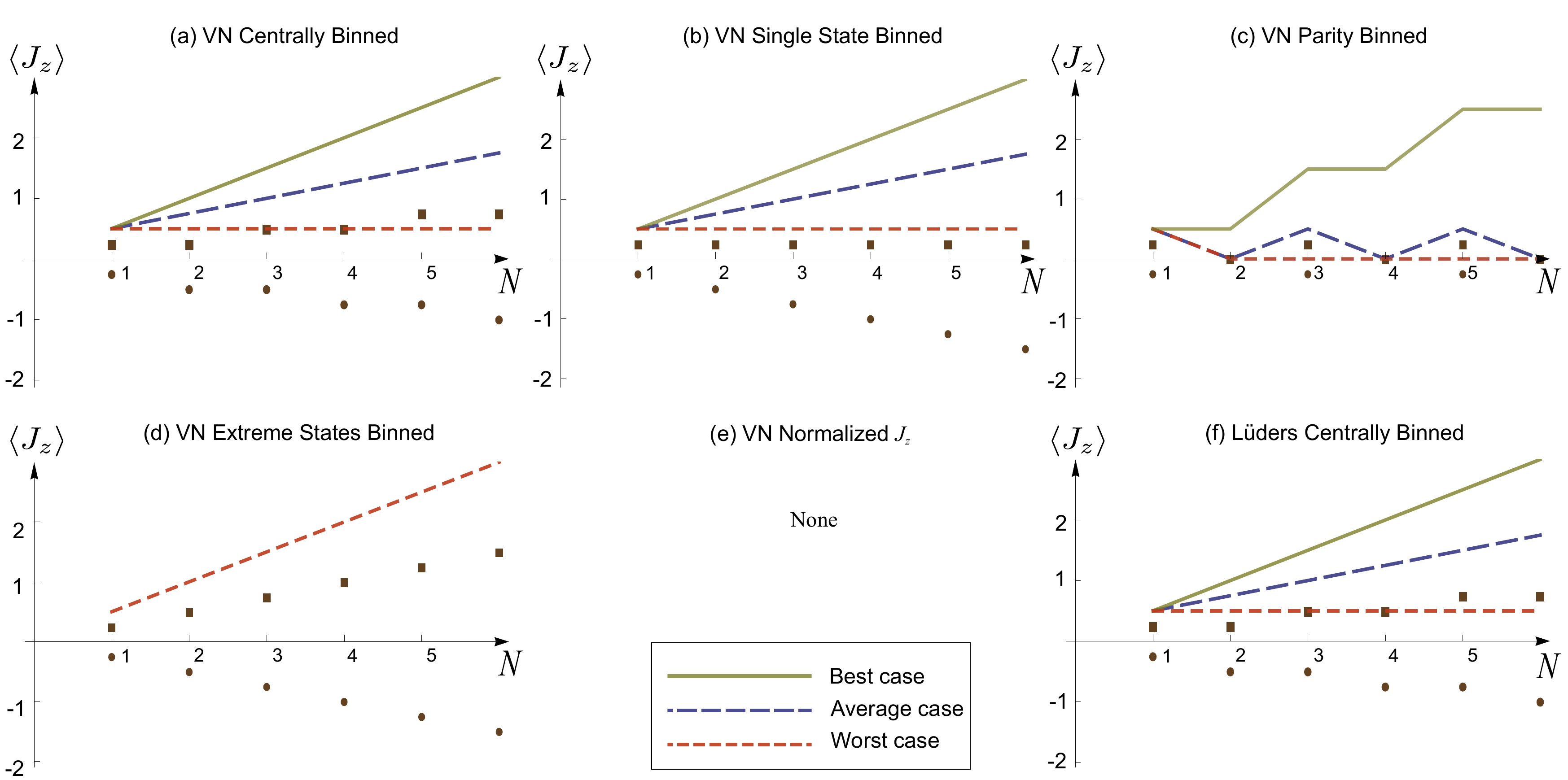}
\caption{\label{grid} (Color online) The best case (green solid line), worst case (red dashed) and linear average (blue dashed ) disconnectivites of each measurement scheme. The expectation values of the different manifolds contributing to the the average case are shown as well, with filled squares for the + manifold, and circles for the - manifold.}
\end{figure*}

\section{Macroscopicity\label{macroscopicity}}

Finally, to understand whether the differing schemes really reflect the macroscopic nature of the ensemble, we adopt Leggett's disconnectivity measure \cite{Leggett1985,Leggett80}:
\bea
\Delta\equiv \mathbb{E}_+-\mathbb{E}_-,
\eea
where $\mathbb{E}_\pm=\langle \psi_\pm|J_z| \psi_\pm\rangle$ is the expectation value of our chosen extensive variable $J_z$ in either of two states $|\psi_\pm\rangle$. We know that violation of the LGI is evidence of quantum coherence between $|\psi_+\rangle$ and $|\psi_-\rangle$.  However, for our chosen binning schemes, each of theses states generally has an internal structure; thus it is not immediately obvious how we should calculate the correct $\Delta$ that applies in each case. One possibility that we consider here is to look at the possible distributions over said internal structure. As such, we define $\Delta_{\text{best}},\Delta_{\text{worst}},\Delta_{\text{av}}$, as the largest, smallest, and average disconnectivity measures, respectively. The average measure corresponds to a uniform weighting over the internal energy levels. For the moment, we shall ignore scheme (e) because it does not define only two states. Further, scheme (f) will not be explicitly discussed because it is equivalent to scheme (a) vis-\`{a}-vis macroscopicity. The results we define below are plotted in \figref{grid}.

\subsection{Best case for macroscopicity}
By inspecting \figref{fig:schemes}, one can see that for schemes (a,b,d), one has $\Delta_{\text{best}}=N$. Scheme (c) can also reach this behaviour when $N$ is odd; otherwise, there is a small correction to $\Delta^{\text{even}}_{\text{best}}=N-1$ .
\subsection{Average case for macroscopicity}
We use
\bea
\Delta_{\text{av}}\equiv\left\{\frac{1}{M_+}\sum_{m\in |\psi_+\rangle} -\frac{1}{M_-}\sum_{m\in |\psi_-\rangle}\right\}\langle m|J_z|m\rangle,
\eea
where the sums run over the $M_\pm$ eigenstates of  $J_z$ in the $|\psi_\pm\rangle$ manifold. The relation $\langle m|J_z|m\rangle=m-\frac{N}{2}$ reveals
\begin{align}
\Delta_{\text{av}}^{(a)}=
\begin{cases}
\frac{1}{2}(N+1) & N \textrm{ odd}\\
\frac{1}{2}N & N\textrm{ even}
\end{cases}
\end{align}
\begin{align}
\Delta_{\text{av}}^{(b)}=\frac{1}{2}(N+1)
\end{align}
\begin{align}
\Delta_{\text{av}}^{(c)}=
\begin{cases}
1 & N \textrm{ odd}\\
0 & N\textrm{ even}
\end{cases}
\end{align}
\begin{align}
\Delta_{\text{av}}^{(d)}=N.
\end{align}
It is interesting to note that (c) has the worst average performance, with an extensive difference which is either null or unity. So increasing the size of the ensemble would not show higher degrees of macroscopicity (on average) in this case.
\subsection{Worst case for macroscopicity}
As is often true, the most important quantities are in the worst-case scenarios. Schemes (a) and (b) have a worst-case extensive difference of $\Delta_{\text{worst}}^{(a)}=\Delta_{\text{worst}}^{(b)}=1$: we cannot exclude the possibility that coherence existed only between neighbouring states on the $J_z$ ladder. Scheme (c) is even worse: $\Delta_{\text{worst}}^{(c)}=0$ for $N>1$, since we may have coherence only between degenerate manifolds. For scheme (d), however, we cannot deny coherence with extensive difference $\Delta_{\text{worst}}^{(d)}=N$, which is potentially \emph{macroscopically distinct}.

\section{Experimental Realizations \label{experiments}}

	As mentioned in the Introduction, there is a great range of experimental systems with which it would be feasible to test the results we have discussed in this work. As we have shown, the requirements in terms of  noise are modest, as the maximum of the LGI violation tend to occur at very short times. We discuss some approaches to how to perform the measurements in \appref{measurements}.\\

	For typical ensembles of flux qubits, with frequencies in the range of GHz, coherence times of between $\unit[1]{\mu s}$ and $\unit[40]{\mu s}$ for single qubits \cite{Steffen2010, Corcoles2011,Yan2015,Knee2016} have been observed. For an ensemble of $6$ qubits coupled to a 3D resonator, a coherence time of  $2$--$\unit[8]{\mu s}$ was observed \cite{Stern2014}. So far ensembles of $20$ qubits have been fabricated \cite{macha}, but ensembles of upto $5000$ qubits, coupled to a common cavity for readout, seem feasible. %With the recent reports of experimentalists achieving the strong ($g/\omega \approx$ 0.1) \cite{Chen2016} and deep-strong ($g/\omega \approx$ 0.72-1.34) coupling regimes \cite{Yoshihara2016} between such qubits and a cavity, fast manipulation and readout of a large ensemble seems feasible in the near future.
 In addition, as each flux qubit itself can arguably be considered macroscopic in nature, a large ensemble of similar devices would be more macroscopic than many other possible realizations. \\

	Other possible systems with which to observe macroscopic LGI violations include atomic spin ensembles. Such ensembles can be coupled to  superconducting resonator cavities, SQUIDS, or even ancilla flux qubits, for manipulation and readout purposes. For example, very recently, Bushev \textit{et al.}~\cite{Bushev2011} revealed the ESR spectroscopy of a spin-cavity system by coupling  Er$^{3+}$ doped Y$_2$SiO$_5$ crystal with a high-Q superconducting resonator. %With a nominal doping of approximately 0.02 \% on a crystal of (1 $\times$ 1.5 $\times$ 3) mm$^3$ size,
 They were able to couple approximately $10^{15}$ spins to the resonator. With varying combinations of doping and temperature, a coherence time of $\unit[20]{m s}$ has reportedly been achieved \cite{B??ttger2006}.  \\

	Similarly, for Al$_2$O$_3$ crystal doped with Cr$^{3+}$, Schuster \textit{et al.}~\cite{Schuster2010} coupled approximately $10^{13}$ spins to a cavity. Similar setups can also be engineered using NV centers in diamond, with a nitrogen density of $\unit[10^{15}]{cm^{-3}}$ and an NV center density $\unit[10^{12}]{cm^{-3}}$. The coherence time was observed to be up to $\unit[0.6]{s}$ at $\unit[77]{K}$ and $\unit[3.3]{ms}$ at room temperature \cite{Bar-Gill2013}. Very recently, Tyryshkin \textit{et al.} \cite{Tyryshkin2011} reported a maximum coherence of up to $\unit[2]{s}$ using silicon doped with a $\unit[50]{ppm}$ concentration of its isotope $^{29}$Si. With rapid developments in fabrication and coherent control of these spin-based systems, it seems possible that the Leggett-Garg violations and the concepts of macroscopicity can be tested in the near future with such large ensembles.

\section{Discussion}

Our results suggest that in designing experiments for observing LGI violations with large ensembles of qubits, one must choose between observing a robust large violation, like with scheme (a), or a harder but more macroscopic measurement, with an extreme-states measurement, as in scheme (d). In this way our results begin to show how an experimentalist, tasked with demonstrating a superposition of macroscopically distinct states in the laboratory, might go about exploiting the tradeoffs between the degree of macroscopicity and constraints on time and the nature of the measurement process in qubit ensembles (both naturally occurring and engineered). Due to the ubiquitous and unavoidable nature of noise in such ensembles, the conclusions we draw concerning the optimal measurement scheme become all the more relevant.

Lastly, given the realistic parameters we have used in our model, we predict that {\em a violation of the Leggett-Garg inequality in an ensemble of between $N=10^8$  and $10^{13}$ NV centers, or several thousand flux qubits, should be possible in the near future}. Readout times of flux qubits have been performed on the time scale of $\unit[140]{ns}$ with $99.8$\% fidelity \cite{Evan2014}, which can be improved with alternative measurement techniques \cite{Didier2015} (see section \appref{measurements}). Because of the robustness to noise, scheme (a) is our recommendation. However, a more ambitious experiment using scheme (d) should also be possible in some systems, as long as the collective and individual qubit dissipation rates could be reduced. Then, coherence between states of unprecedented macroscopic distinctness could be possible.

\acknowledgments

We thank Clive Emary and Yuichiro Matsuzaki for helpful feedback.  This work is partially supported by the RIKEN iTHES Project, the MURI Center for Dynamic Magneto-Optics via the AFOSR award number FA9550-14-1-0040, the IMPACT program of JST, and a Grant-in-Aid for Scientific Research (A). NL is partially supported by the FY2015 Incentive Research Project. AFK acknowledges support from a JSPS Postdoctoral Fellowship for Overseas Researchers. GCK was supported by the Royal Commission for the Exhibition of 1851.  KD was supported by the RIKEN-IPA program.

\appendix

\section{Analytical results \label{AppA}}

In this Appendix, we derive expressions for the LG parameter $K$ for our measurement schemes. We begin by rewriting the correlation functions in $K$ as sums of matrix elements for the time-evolution operator, then calculate these matrix elements for the Hamiltonian of our system, and finally perform further simplifications where possible for the different measurement schemes.  Note that in the following we set $\hbar = 1$ for notational simplicity.

\subsection{Correlation functions}

We consider the LG parameter from \eqref{eq:DefK}, repeated here for convenience,
\be
K = C_{21} + C_{32} - C_{31}.
\ee
Here, the correlation functions are
\be
C_{ba} = \expec{Q(t_b)Q(t_a)},
\ee
where $t_b > t_a$ and $Q$ is a measurement result that can take the values $\pm1$, apart from the normalized $J_z$ measurement, where $Q$ takes values in the range $\{-1,1\}$.

We treat our ensemble of $N$ qubits as a large spin of magnitude $j=\frac{N}{2}$. Starting in the state $\rho(0) = \ketbra{j}{j}$ (we are only writing the quantum number $m$ in the kets here), performing measurements with results $q_m$ and projection operators $\Pi_m = \ketbra{m}{m}$, and writing the time evolution between measurements in \eqref{eq:CorrelationFunction} as a unitary evolution,
 $\rho(t) = \mathcal{U}(t- t_0) \rho (t_0) \equiv U(t,t_0) \rho (t_0) U^\dag(t,t_0)$ , we obtain
\begin{widetext}
\bea
\expec{Q(t_b)Q(t_a)} &=& \sum_{n,m} q_n q_m \tr{\Pi_m U(t_b,t_a) \Pi_n U(t_a,0) \rho(0) U^\dag(t_a,0) \Pi_n U^\dag(t_b,t_a) \Pi_m} \nn\\
%&=& \sum_{n,m} q_n q_m \tr{\ketbra{m}{m} U(t_b,t_a) \ketbra{n}{n} U(t_a,0) \ketbra{j}{j} U^\dag(t_a,0) \ketbra{n}{n} U^\dag(t_b,t_a) \ketbra{m}{m}} \nn\\
%&=& \sum_{n,m,k} q_n q_m \braket{k}{m} \brakket{m}{U(t_b,t_a)}{n} \brakket{n}{U(t_a,0)}{j} \brakket{j}{U^\dag(t_a,0)}{n} \brakket{n}{U^\dag(t_b,t_a)}{m} \braket{m}{k} \nn\\
&=& \sum_{n,m} q_n q_m \brakket{m}{U(t_b,t_a)}{n} \brakket{n}{U(t_a,0)}{j} \brakket{j}{U^\dag(t_a,0)}{n} \brakket{n}{U^\dag(t_b,t_a)}{m} \nn\\
&=& \sum_{n,m} q_n q_m \abssq{\brakket{m}{U(t_b,t_a)}{n}} \abssq{\brakket{n}{U(t_a,0)}{j}},
\label{eq:CorrelationFunctionFormula}
\eea
\end{widetext}
where we used $\brakket{a}{O}{b} = \brakket{b}{O^\dag}{a}^\dag$ in the last step.

We consider the case $t_1 = 0$, $t_2 = \tau$, and $t_3 = 2\tau$. Then, with the abbreviated notation $U(t,t_0) = U(t-t_0)$, we obtain from \eqref{eq:CorrelationFunctionFormula} the three correlation functions
%\begin{widetext}
\bea
C_{21} &=& %\sum_{n,m} q_n q_m \abssq{\brakket{m}{U(\tau)}{n}} \abssq{\braket{n}{j}} =
q_j \sum_{m} q_m \abssq{\brakket{m}{U(\tau)}{j}}, \label{C21Sum}\\
C_{31} &=& %\sum_{n,m} q_n q_m \abssq{\brakket{m}{U(2\tau)}{n}} \abssq{\braket{n}{j}} =
q_j \sum_{m} q_m \abssq{\brakket{m}{U(2\tau)}{j}},  \label{C31Sum} \\
C_{32} &=& \sum_{n,m} q_n q_m \abssq{\brakket{m}{U(\tau)}{n}} \abssq{\brakket{n}{U(\tau)}{j}}.  \label{C32Sum}
\eea
%\end{widetext}
Depending on the choice of $q_m$ and $\tau$, these expressions may be simplified further.

\subsection{Matrix elements}
\label{sec:MatrixElements}

From \eqref{eq:Hamiltonian}, we have that our giant spin evolves under the Hamiltonian
\be
H = \Omega J_x,
\ee
and the time evolution operator is thus
\be
U(\tau) = \exp{(-iJ_x\Omega\tau)},
\ee
which represents a rotation of the spin. The matrix elements for general spin rotations, parameterized by the Euler angles $\alpha, \beta, \gamma$, is given by the Wigner D-matrix \cite{Sakurai1994}
%\begin{widetext}
\bea
D^{(j)}_{m,m'} (\alpha,\beta,\gamma) &=& \brakket{j,m'}{e^{-iJ_z\alpha} e^{-iJ_y\beta} e^{-iJ_z\gamma}}{j,m} \nn\\
&=& e^{-i (m'\alpha + m \gamma)} \brakket{j,m'}{e^{-iJ_y\beta}}{j,m} \nn\\
&=&  e^{-i (m'\alpha + m \gamma)} d^{(j)}_{m,m'} (\beta),
\eea
%\end{widetext}
where the small $d$-matrix is
\begin{widetext}
\bea
d^{(j)}_{m,m'} (\beta) &=& \sum_k \left(-1\right)^{k-m+m'} \frac{\sqrt{(j+m)! (j-m)! (j+m')! (j-m')!}}{(j+m-k)! k! (j-k-m')! (k-m+m')!}% \nn\\
%&&\times
\left[\cos\left(\frac{\beta}{2}\right)\right]^{2j-2k+m-m'} \left[\sin\left(\frac{\beta}{2}\right)\right]^{2k-m+m'}. \nn\\
\eea
\end{widetext}
Here, the sum is over all $k$ such that none of the factorials in the denominator are evaluated for negative numbers.

In our case, we have a rotation around the $x$ axis by an angle $\Omega\tau$. This can be decomposed into rotations around the $z$ axis and a rotation around the $y$ axis by the same angle $\Omega\tau$. Since we only need the absolute value squared of the matrix element $D^{(j)}_{m,m'} (\alpha,\beta,\gamma)$ to calculate the correlation functions, it suffices to evaluate
\begin{widetext}
\bea
&&\abssq{\brakket{m}{\exp{(-iJ_x\Omega\tau)}}{n}} = \abssq{d^{(j)}_{n,m} (\Omega\tau)} \nn\\
=&& \abssq{\sum_k \left(-1\right)^{k} \frac{\sqrt{(j+n)! (j-n)! (j+m)! (j-m)!}}{(j+n-k)! k! (j-k-m)! (k-n+m)!} \left[\cos\left(\frac{\Omega\tau}{2}\right)\right]^{2j-2k+n-m} \left[\sin\left(\frac{\Omega\tau}{2}\right)\right]^{2k-n+m}}.
\eea
\end{widetext}

The sum over $k$ simplifies to fewer terms in a few special cases, where $n$ and/or $m$ equals $\pm j$. Some of these cases are relevant for the different measurement schemes we consider, so we calculate them below. However, first of all, we note the restrictions on $k$ in the general expression above: from the terms in the denominator we derive the conditions $k \leq j+n$, $k \geq 0$, $k \leq j-m$, and $k \geq n-m$, which means that the sum goes over all $k$ in the interval $\max(0,n-m) \leq k \leq \min(j-m,j+n)$.

For the case $n=j$, the restrictions on $k$ means that only $k=j-m$ contributes to the sum. We obtain
%\begin{widetext}
\bea
&&\abssq{\brakket{m}{\exp{(-iJ_x\Omega\tau)}}{j}} \nn\\ %&=& \abssq{\frac{\sqrt{(2j)! 0! (j+m)! (j-m)!}}{(j+m)! (j-m)! 0! 0!} \left(\cos\left(\frac{\Omega\tau}{2}\right)\right)^{j+m} \left(\sin\left(\frac{\Omega\tau}{2}\right)\right)^{j-m}} \nn\\
%&=& \frac{(2j)!}{(j+m)! (j-m)!} \left(\cos\left(\frac{\Omega\tau}{2}\right)\right)^{2(j+m)} \left(\sin\left(\frac{\Omega\tau}{2}\right)\right)^{2(j-m)} \nn\\
&=& \binom{2j}{j+m} \left[\cos\left(\frac{\Omega\tau}{2}\right)\right]^{2(j+m)} \left[\sin\left(\frac{\Omega\tau}{2}\right)\right]^{2(j-m)}, \nn\\
\eea
%\end{widetext}
where we used $\binom{a}{b} = a!/[b!(a-b)!]$.
%Here, we make the observation that at times $\tau$ such that $\cos\left(\frac{\Omega\tau}{2}\right) = 0$, only $m=-j$ gives a nonzero matrix element (since $0^0 = 1$), and, in the same way, only $m=j$ gives a nonzero matrix element when $\sin\left(\frac{\Omega\tau}{2}\right) = 0$. This is consistent with the picture of $\pi$ and $2\pi$ rotations of the large spin.

From the above, we can immediately compute the even more special cases $n=j, m=\pm j$:
%\begin{widetext}
\bea
\abssq{\brakket{j}{\exp{(-iJ_x\Omega\tau)}}{j}} &=& %\binom{2j}{2j} \left(\cos\left(\frac{\Omega\tau}{2}\right)\right)^{2(j+j)} \left(\sin\left(\frac{\Omega\tau}{2}\right)\right)^{2(j-j)} =
\left[\cos\left(\frac{\Omega\tau}{2}\right)\right]^{4j}, \\
\abssq{\brakket{-j}{\exp{(-iJ_x\Omega\tau)}}{j}} &=& %\binom{2j}{0} \left(\cos\left(\frac{\Omega\tau}{2}\right)\right)^{2(j-j)} \left(\sin\left(\frac{\Omega\tau}{2}\right)\right)^{2(j+j)} =
\left[\sin\left(\frac{\Omega\tau}{2}\right)\right]^{4j}.
\eea
%\end{widetext}

Finally, we also consider the case $n=-j$, for which we see that only $k=0$ contributes to the sum. We thus get
%\begin{widetext}
\bea
&& \abssq{\brakket{m}{\exp{(-iJ_x\Omega\tau)}}{-j}} \nn\\
%&=& \abssq{\frac{\sqrt{0! (2j)! (j+m)! (j-m)!}}{0! 0! (j-m)! (j+m)!} \left(\cos\left(\frac{\Omega\tau}{2}\right)\right)^{j-m} \left(\sin\left(\frac{\Omega\tau}{2}\right)\right)^{j+m}} \nn\\
&=& \binom{2j}{j+m} \left[\cos\left(\frac{\Omega\tau}{2}\right)\right]^{2(j-m)} \left[\sin\left(\frac{\Omega\tau}{2}\right)\right]^{2(j+m)}, \nn\\
\eea
%\end{widetext}
and in the more specialized cases with $m=\pm j$, the result is
\bea
\abssq{\brakket{j}{\exp{(-iJ_x\Omega\tau)}}{-j}} &=& \left[\sin\left(\frac{\Omega\tau}{2}\right)\right]^{4j}, \\
\abssq{\brakket{-j}{\exp{(-iJ_x\Omega\tau)}}{-j}} &=& \left[\cos\left(\frac{\Omega\tau}{2}\right)\right]^{4j}.
\eea

\subsection{Evaluating $K$ for the different measurement schemes}

\subsubsection{von Neumann centrally binned}

For this scheme, we use $q_{m\geq 0} = +1$ and $q_{m<0} = -1$. In this case, inserting the matrix elements calculated in \secref{sec:MatrixElements} into Eqs.~(\ref{C21Sum})-(\ref{C32Sum}) gives
\begin{widetext}
\bea
C_{21} &=& %q_j \sum_{m} q_m \abssq{\brakket{m}{U(\tau)}{j}} =
\sum_{m} q_m \binom{2j}{j+m} \left[\cos\left(\frac{\Omega\tau}{2}\right)\right]^{2(j+m)} \left[\sin\left(\frac{\Omega\tau}{2}\right)\right]^{2(j-m)}, \\
C_{31} &=& %q_j \sum_{m} q_m \abssq{\brakket{m}{U(2\tau)}{j}} =
\sum_{m} q_m \binom{2j}{j+m} [\cos(\Omega\tau)]^{2(j+m)} [\sin(\Omega\tau)]^{2(j-m)},
\eea

\bea
C_{32} &=& %\sum_{n,m} q_n q_m \abssq{\brakket{m}{U(\tau)}{n}} \abssq{\brakket{n}{U(\tau)}{j}} \nn\\
\sum_{n,m} q_n q_m \binom{2j}{j+n} \left[\cos\left(\frac{\Omega\tau}{2}\right)\right]^{2(j+n)} \left[\sin\left(\frac{\Omega\tau}{2}\right)\right]^{2(j-n)} \nn\\
&&\times \abssq{\sum_{k=\max(0,n-m)}^{\min(j-m,j+n)} \left(-1\right)^{k} \frac{\sqrt{(j+n)! (j-n)! (j+m)! (j-m)!}}{(j+n-k)! k! (j-k-m)! (k-n+m)!} \left[\cos\left(\frac{\Omega\tau}{2}\right)\right]^{2j-2k+n-m} \left[\sin\left(\frac{\Omega\tau}{2}\right)\right]^{2k-n+m}}. \nn\\
\eea
%\end{widetext}
and the Leggett--Garg parameter $K$ is thus
%\begin{widetext}
\bea
K &=& %C_{21} + C_{32} - C_{31} \nn\\
%&=&
\sum_{m} q_m \binom{2j}{j+m} \Bigg\{ \left[\cos\left(\frac{\Omega\tau}{2}\right)\right]^{2(j+m)} \left[\sin\left(\frac{\Omega\tau}{2}\right)\right]^{2(j-m)} \nn\\
&&\times \Bigg(1 + \sum_n q_n \Bigg| \sum_{k=\max(0,m-n)}^{\min(j-n,j+m)} \left(-1\right)^{k} \frac{\sqrt{(j+m)! (j-m)! (j+n)! (j-n)!}}{(j+m-k)! k! (j-k-n)! (k-m+n)!} \nn\\
&&\times \left[\cos\left(\frac{\Omega\tau}{2}\right)\right]^{2j-2k+m-n} \left[\sin\left(\frac{\Omega\tau}{2}\right)\right]^{2k-m+n}\Bigg|^2 \Bigg) - [\cos(\Omega\tau)]^{2(j+m)} [\sin(\Omega\tau)]^{2(j-m)} \Bigg\}. \label{eq:KDichotomic}
\eea
\end{widetext}
%\begin{figure}\centering
%\includegraphics[width=0.8\linewidth]{DichotomicPlot.eps}
%\caption{The value of the Leggett-Garg parameter $K$ for $N=1,3,10,50$ for the dichotomic measurement. The dotted light blue line marks $K=1$, the border for the classical world. \label{fig:Dichotomic}}
%\end{figure}
%
The value of $K$ is plotted as a function of $\Omega\tau$ in \figref{fig:KvsTime}a. The maximum for $K$ seems to be reached around $\Omega\tau = \pi/4$. Plugging this value into \eqref{eq:KDichotomic} unfortunately does not give any significant simplifications.

\subsubsection{von Neumann single state binning}

%Budroni and Emary \cite{Budroni2014} used yet another measurement scheme, $q_m = 1 - 2\delta_{m,-j}$.
This measurement scheme, where $q_m = 1 - 2\delta_{m,-j}$, was used by Budroni and Emary \cite{Budroni2014}.
The formula we have for $K$ in \eqref{eq:KDichotomic} applies here too and is used to plot $K$ as a function of $\Omega\tau$ for this scheme in \figref{fig:KvsTime}b. In this case, the maximum is reached around $\Omega\tau = \pi/2$, which allows for some simplifications. Furthermore, this form of $q_m$ allows one to simplify all the sums using the resolution of identity, and in the end one only needs the matrix elements where $n$ and $m$ are $\pm j$.  As shown in the appendix of Ref.~\cite{Budroni2014}, this leads to a simple analytical formula for the maximum value of $K$ for large spins:
\be
K_{max} = 3 - \sqrt{\frac{2}{\pi j}},
\ee
which approaches 3 when $j \rightarrow \infty$.

\subsubsection{von Neumann parity binning}

For parity binning, we use $q_m = +1$ for $m = j, j-2, j-4, \ldots$ and $q_m = -1$ for $m = j-1, j-3, j-5, \ldots$. The result from \eqref{eq:KDichotomic} applies for this scheme as well and is used to plot $K$ as a function of $\Omega\tau$ in \figref{fig:KvsTime}c. The maximum for $K$ seems to be reached close to $\Omega\tau = 0$ for large $N$. Even if we can expand the trigonometric parts of $K$ around this point, the large sums still remain and further analytical simplifications remain out of reach.

\subsubsection{von Neumann extreme states binning}

In this scheme, all runs of the experiment resulting in $m \neq \pm j$ are discarded. The remaining cases are assigned the measurement results $q_j = +1, q_{-j} = -1$. This considerably simplifies the sums in Eqs.~(\ref{C21Sum})-(\ref{C32Sum}). Using the matrix elements calculated in the previous section, we obtain
%\begin{widetext}
\bea
C_{21} &=& %\sum_{n,m} q_n q_m \abssq{\brakket{m}{U(\tau)}{n}} \abssq{\braket{n}{j}} =
q_j \sum_{m} q_m \abssq{\brakket{m}{U(\tau)}{j}} \nn\\
&=& \abssq{\brakket{j}{U(\tau)}{j}} - \abssq{\brakket{-j}{U(\tau)}{j}} \nn\\
&=& \left[\cos\left(\frac{\Omega\tau}{2}\right)\right]^{4j} - \left[\sin\left(\frac{\Omega\tau}{2}\right)\right]^{4j}.
\eea
%\end{widetext}
$C_{31}$ is simply $C_{21}$ with $\tau$ replaced by $2\tau$:
\bea
C_{31} = [\cos(\Omega\tau)]^{4j} - [\sin(\Omega\tau)]^{4j}.
\eea
%
%$C_{32}$, there is a double sum, but this does not really complicate things here:
The calculation for $C_{32}$ is similar to that for $C_{21}$ and gives
%\begin{widetext}
\bea
%C_{32} &=& \sum_{n,m} q_n q_m \abssq{\brakket{m}{U(\tau)}{n}} \abssq{\brakket{n}{U(\tau)}{j}} \nn\\
%&=& \abssq{\brakket{j}{U(\tau)}{j}}\abssq{\brakket{j}{U(\tau)}{j}} - \abssq{\brakket{-j}{U(\tau)}{j}}\abssq{\brakket{j}{U(\tau)}{j}} \nn\\
%&&- \abssq{\brakket{j}{U(\tau)}{-j}}\abssq{\brakket{-j}{U(\tau)}{j}} + \abssq{\brakket{-j}{U(\tau)}{-j}}\abssq{\brakket{-j}{U(\tau)}{j}} \nn\\
%&=& \left(\cos\left(\frac{\Omega\tau}{2}\right)\right)^{8j} - \left(\sin\left(\frac{\Omega\tau}{2}\right)\right)^{4j} \left(\cos\left(\frac{\Omega\tau}{2}\right)\right)^{4j} - \left(\sin\left(\frac{\Omega\tau}{2}\right)\right)^{8j} + \left(\cos\left(\frac{\Omega\tau}{2}\right)\right)^{4j} \left(\sin\left(\frac{\Omega\tau}{2}\right)\right)^{4j} \nn\\
C_{32} &=& \left[\cos\left(\frac{\Omega\tau}{2}\right)\right]^{8j} - \left[\sin\left(\frac{\Omega\tau}{2}\right)\right]^{8j}.
\eea
%\end{widetext}
Thus, the Leggett-Garg parameter $K$ becomes
%\begin{widetext}
\bea
%K &=& C_{21} + C_{32} - C_{31} \nn\\
K &=& \left[\cos\left(\frac{\Omega\tau}{2}\right)\right]^{4j} - \left[\sin\left(\frac{\Omega\tau}{2}\right)\right]^{4j} \nn\\
&&+ \left[\cos\left(\frac{\Omega\tau}{2}\right)\right]^{8j} - \left[\sin\left(\frac{\Omega\tau}{2}\right)\right]^{8j} \nn\\
&&- [\cos(\Omega\tau)]^{4j} + [\sin(\Omega\tau)]^{4j}.
\eea
%\end{widetext}

%\begin{figure}\centering
%\includegraphics[width=0.8\linewidth]{ExtremeStatesPlot.eps}
%\caption{The value of the Leggett-Garg parameter $K$ for $N=1,3,10,50$ for the extreme states measurement. The dotted light blue line marks $K=1$, the border for the classical world. \label{fig:ExtremeStates}}
%\end{figure}

The value of $K$ is plotted as a function of $\Omega\tau$ for this scheme in \figref{fig:KvsTime}d. We note that as $N$ increases, the maximum of $K$ decreases and occurs close to $\Omega\tau = 0$. To find the asymptotic behaviour of $K_{\mathrm{max}}$, we can try to expand $K$ for small values of $\Omega\tau$. However, the terms of order $2n$ in $\Omega\tau$ in that expansion have coefficients proportional to $j^n$, which prevents us from finding an approximate asymptote when $j \rightarrow \infty$. Fortunately, the simple formula for $K$ allows for numerical investigations for very large $j$, which indicate that $K_{\mathrm{max}} \rightarrow 1.055$ in the limit of many qubits.

\subsubsection{von Neumann normalized $J_z$ measurement}

For the normalized $J_z$ measurement, we use $q_m = m/j$. Just like for the VN single state binning scheme and the parity binning scheme, we can use \eqref{eq:KDichotomic} with the new definition of $q_m$.
%The result is
%\begin{widetext}
%\bea
%K &=& \sum_{m} \frac{q_m}{j} \binom{2j}{j+m} \Bigg[ \left(\cos\left(\frac{\Omega\tau}{2}\right)\right)^{2(j+m)} \left(\sin\left(\frac{\Omega\tau}{2}\right)\right)^{2(j-m)} \nn\\
%&&\times \Bigg(1 + \sum_n \frac{q_n}{j} \Bigg| \sum_{k=\max(0,m-n)}^{\min(j-n,j+m)} \left(-1\right)^{k} \frac{\sqrt{(j+m)! (j-m)! (j+n)! (j-n)!}}{(j+m-k)! k! (j-k-n)! (k-m+n)!} \nn\\
%&&\times \left(\cos\left(\frac{\Omega\tau}{2}\right)\right)^{2j-2k+m-n} \left(\sin\left(\frac{\Omega\tau}{2}\right)\right)^{2k-m+n}\Bigg|^2 \Bigg) - (\cos(\Omega\tau))^{2(j+m)} (\sin(\Omega\tau))^{2(j-m)} \Bigg]. \label{eq:KNormalizedJz}
%\eea
%\end{widetext}
The value of $K$ is plotted as a function of $\Omega\tau$ in \figref{fig:KvsTime}e. The maximum for $K$ seems to be reached somewhere between $\Omega\tau = \pi/8$ and $\Omega\tau = \pi/4$ for large $N$. Further analytical simplifications to find the asymptotic behaviour of $K_{\mathrm{max}}$ are not possible here.
%Unfortunately, it is not at $\Omega\tau = \pi/4$ (here, $K=1$), so again further analytical simplifications seems out of reach.
%
%\begin{figure}\centering
%\includegraphics[width=0.8\linewidth]{NormalizedJzPlot.eps}
%\caption{The value of the Leggett-Garg parameter $K$ for $N=1,3,10,50$ for the normalized $J_z$ measurement. The dotted light blue line marks $K=1$, the border for the classical world. \label{fig:NormalizedJz}}
%\end{figure}
%

\subsubsection{Dichotomic measurement with the L\"uders state update rule}

This scheme is different from the rest in that the system is not projected onto a single-spin eigenstate after a measurement, but onto a large superposition of spin eigenstates [see \eqref{eq:LuderPostMsmt}]. Thus, if we calculate the correlation functions as in \eqref{eq:CorrelationFunctionFormula}, we are left with a large number of sums, which are not amenable to analytical simplifications.

\section{Cavity-based measurements \label{measurements}}

One of the advantages of the large ensembles of flux qubits or spin ensembles we outlined in \secref{experiments} is that the collective large-spin degree of freedom can be read out with a range of well-developed techniques, typically used for the purposes of quantum information protocols or quantum simulation. For example, one may couple the ensemble to a common microwave transmission line cavity mode \cite{Evan2014,Didier2015}, leading  to a dispersive interaction between the ensemble and measurement-cavity system, similar to that derived for a large-spin in Ref.~\cite{disp}.

However, this is not ideal for our purposes, as the dispersive interaction term is only the lowest order in perturbation theory, and higher-order terms would constitute an invasive or clumsy measurement of the cavity onto the ensemble, which we wish to avoid. One should also note that, in making the dispersive transformation discussed in Ref.~\cite{disp}, the ensemble's collective interaction with the cavity  creates an additional spin-spin coupling term, mediated by virtual-excitation exchange with the cavity, which gives rise to a spin squeezing $ J_+J_-$ term in the large-spin basis.  This, along with a cavity-induced superradiant decay of the large spin, similar to the collective dissipation term $\Gamma_L$ we used in the examples earlier, constitutes an additional unwanted backaction during the measurement process.

If one wished to proceed with this approach in any case, one could probe the cavity with a weak field at a single frequency, and thus check whether the ensemble is in just one state, or not that state, directly realizing a hybrid of scheme (b) with  the L\"uders post-measurement rule, as in scheme (f).  Alternatively, one can observe the quadrature phase shift of the cavity field, whose phase and magnitude depend on the $J_z$ value of the ensemble. Both approaches may become more difficult as $N$ is increased, however, as the signal to noise (SNR) ratio diminishes (the quadrature displacements of each possible outcome become difficult to distinguish).  This would require a decrease  in the cavity broadening as $N$ increases to maintain the same SNR.  However, this is further restricted as both approaches require a readout of the cavity faster than the ensemble decay time, and without overly populating the cavity itself, the combination of which limits how small the cavity broadening can be.

As an alternative to this dispersive-readout approach one could engineer a time-dependent longitudinal coupling to the cavity and realize fast measurement of the ensemble without either undue disturbance, unwanted spin-spin couplings,  or collective superradiance \cite{Didier2015}.  In addition, the speed of the normal dispersive readout scheme is limited by the perturbative nature of the dispersive interaction; as mentioned above, if the coupling is made stronger, or the number of photons in the cavity is too high, for example, higher-order contributions can lead to unwanted excitation exchange between cavity and ensemble, lessening the impact of the observation of any violation of the LGI. In this longitudinal scheme (see \cite{Didier2015} for details of the single-spin case), when applied to a multi-level system one must look at {\em both} the direction of the cavity quadrature displacement and the amplitude of that displacement to distinguish the different sub-levels of the large spin. In addition, one must decrease the cavity dissipation as $N$ is increased. However, in this case, one can {\em a priori} perform faster readout, and thus this is less of a concern. Thus, we conclude that, despite the engineering difficulties associated with generating a longitudinal time-dependant coupling, this approach to measuring the ensemble seems superior for our purposes than the normal dispersive approach.

While the above approaches may work well for an ensemble of flux qubits, for NV centers an alternative measurement scheme involves coupling the ensemble to a large SQUID, where large dispersive coupling arises naturally, not as an approximation to a full transverse coupling \cite{Matsuzaki2015}, thus circumventing the issue of unwanted backaction and large collective decay.

\bibstyle{apsrev4-1}
\bibliography{LGReferences,references}
\end{document}